\documentclass[pre,superscriptaddress,twocolumn,floatfix,longbibliography]{revtex4-1}
\usepackage{graphics}
\usepackage{graphicx}
\usepackage{amsmath}
\usepackage{amssymb,xcolor}

\newcommand{\textin}[1]{\mbox{\scriptsize{#1}}}

\definecolor{grisclair}{rgb}{0.6,0.6,0.6}

\newcommand{\beq}{\begin{equation}}
\newcommand{\ee}{\end{equation}}

\begin{document}

\title{The role of charge relaxation in electrified tip streaming}
\author{M. Rubio}
\address{Depto.\ de Ingenier\'{\i}a Energ\'etica y Fluidomec\'anica and\\ 
Instituto de las Tecnolog\'{\i}as Avanzadas de la Producci\'on (ITAP),\\
Universidad de Valladolid, E-47003 Valladolid, Spain}
\author{P. Rodr\'{\i}guez-D\'{\i}az}
\address{Depto.\ de Ingenier\'{\i}a Mec\'anica, Energ\'etica y de los Materiales and\\ 
Instituto de Computaci\'on Cient\'{\i}fica Avanzada (ICCAEx),\\
Universidad de Extremadura, E-06006 Badajoz, Spain}
\author{J. M. L\'opez-Herrera}
\address{Depto.\ de Ingenier\'{\i}a Aeroespacial y Mec\'anica de Fluidos,\\
Universidad de Sevilla, E-41092 Sevilla, Spain}
\author{M. A. Herrada}
\address{Depto.\ de Ingenier\'{\i}a Aeroespacial y Mec\'anica de Fluidos,\\
Universidad de Sevilla, E-41092 Sevilla, Spain}
\author{A. M. Gañ\'an-Calvo}
\address{Departamento de Ingenier\'{\i}a Aeroespacial y Mec\'anica de Fluidos,\\
Universidad de Sevilla, E-41092 Sevilla, Spain}
\author{J. M. Montanero}
\address{Depto.\ de Ingenier\'{\i}a Mec\'anica, Energ\'etica y de los Materiales and\\ 
Instituto de Computaci\'on Cient\'{\i}fica Avanzada (ICCAEx),\\
Universidad de Extremadura, E-06006 Badajoz, Spain}

\begin{abstract}
We study experimentally and numerically the onset of tip streaming in an electrified droplet. The experiments show that, for a sufficiently small dimensionless conductivity, the droplet apex oscillates before ejecting a liquid jet. This effect is caused by the limited charge transfer from the bulk to the interface. This reduces the electrostatic pressure at the droplet's stretching tip, preventing liquid ejection. This reduction of electrostatic pressure is compensated for by the electric shear stress arising during the apex oscillations, which eventually leads to the jet formation. The stability limit calculated from the global stability analysis perfectly agrees with the experimental results. However, this analysis predicts non-oscillatory, non-localized instability in all the cases, suggesting that both the oscillatory behavior and the small local scale characterizing tip streaming arise during the nonlinear droplet deformation.
\end{abstract}

\maketitle

\section{Introduction}

% Charge relaxation
Charge relaxation is known to produce relevant effects on the dynamics of low-conductivity drops and jets under the action of electric fields. For instance, charge relaxation over the surface of a jet in an axial electric field makes small-amplitude disturbances grow in an oscillatory manner, contrary to what happens in the perfect dielectric and perfect conductor case \citep{S71c}. The decrease in the liquid conductivity may produce the radial compression of satellite droplets \citep{NLMFY21}, reduce their size, or even cause their complete elimination \citep{LKYY19}.

% Tip streaming
In his pioneering work, \citet{T64} described experimentally the ejection of very fine jets from the conical points of electrified films. A similar phenomenon has also been observed in the poles of levitated bubbles and droplets subjected to strong electric fields \citep{GK64,DAMHL03,AMDL05,GGRMHDMLG08,V19}. The problem has been numerically solved considering the leaky-dielectric approximation \citep{CSHB13,GMT20}, the charge conservative model \citep{GLRM16}, and electrokinetic effects \citep{PBHD16,MMK19}. Scaling laws for the velocity, diameter, and charge of the first-emitted droplet have been derived and experimentally tested \citep{CSHB13,GLRM16,RSGM21}. The predictions provided by the leaky-dielectric model may differ from those of the two other approximations in this singular problem. In fact, the interface can be created so fast that Ohmic conduction may overestimate the injection of charges from the bulk onto the fresh interface. 

% Beroz
\citet{BHB19} have shown that the apex of a sessile droplet placed between two electrodes rapidly stretches and emits a very thin jet when the liquid volume $\hat{\cal V}$ reaches a critical value. The critical conditions for the droplet bursting are not expected to be affected by the liquid properties (except for the surface tension) because both the velocity field and the inner electric field vanish before the instability arises. For sufficiently large distances between the two electrodes and small enough liquid volumes (gravitational Bond numbers), the critical droplet volume only depends on the characteristic electric field $E_0=V/H$, where $V$ and $H$ are the voltage drop and distance between the electrodes, respectively \citep{BHB19}. 

% This work
The electrified tip streaming phenomenon described above constitutes an ideal candidate for studying electrohydrodynamic phenomena. While previous works have determined the instability limit \citep{BHB19} and analyzed the fluid ejection \citep{GLRM16,RSGM21}, we will focus on the instability mechanism and the droplet tip dynamics preceding the jet formation. Attention will be paid to the role played by charge relaxation due to the small value of the liquid dimensionless conductivity (i.e., the ratio between the capillary time to the electric relaxation time).

\section{Methods}

\subsection{Experiments}
\begin{figure}
\resizebox{0.45\textwidth}{!}{\includegraphics{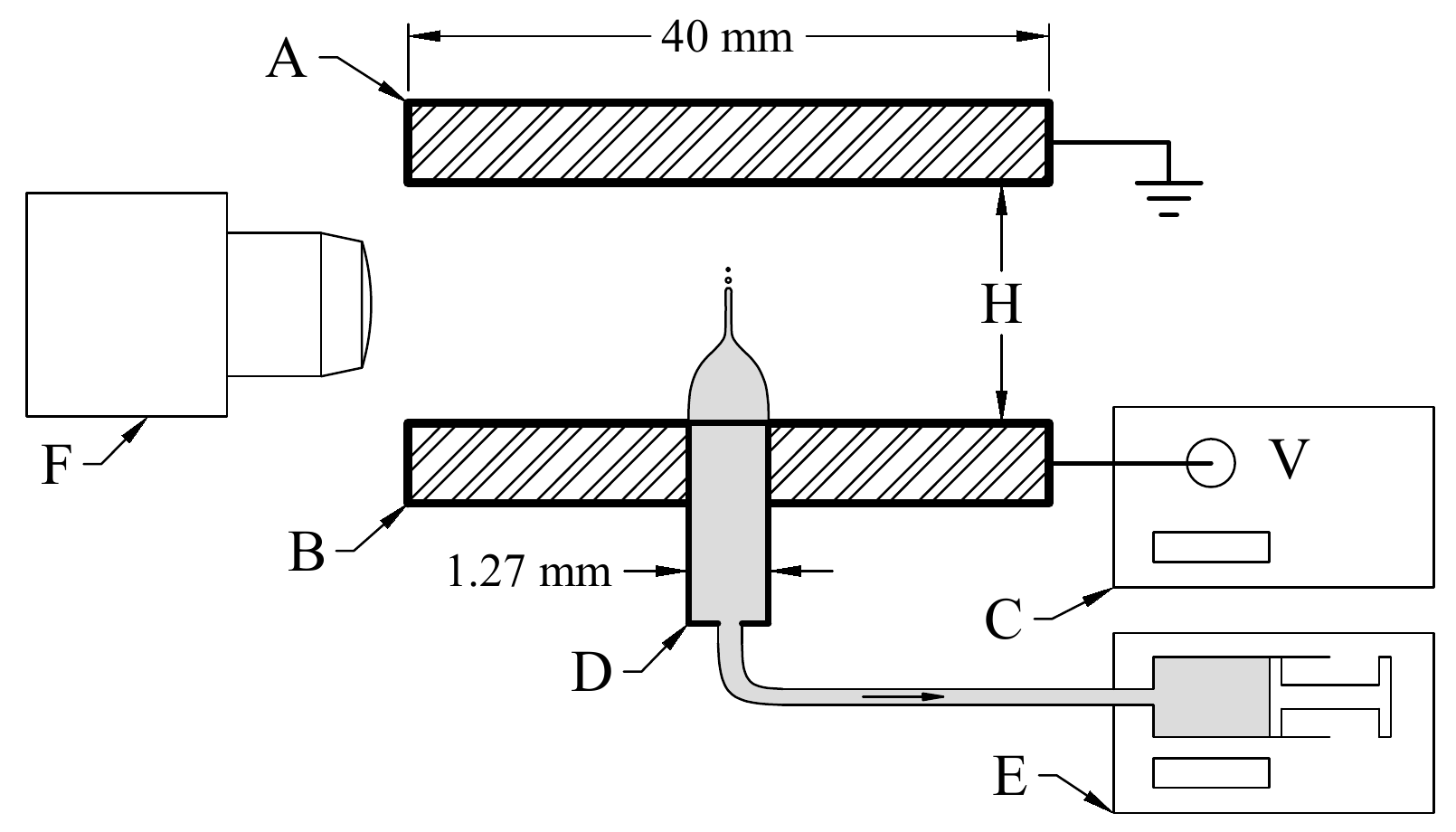}}
\caption{Experimental configuration: upper (A) and lower (B) electrodes, voltage amplifier (C), feeding capillary (D), syringe pump (E), and high-speed camera (F).}
\label{simple}
\end{figure}

In this work, we conducted experiments with the same configuration as that of \citet{BHB19} (Fig.\ \ref{simple}). It consisted of two parallel circular electrodes 40 mm in diameter and made of stainless steel. A cylindrical capillary of radius $R=0.635$ mm was introduced in the center of the lower electrode. This element allowed us to inject the liquid and ensured the anchorage of the triple contact line. The electrodes were separated by a distance $H=3.3$ mm. 

A constant voltage $V=5$ kV was applied to the lower electrode using a voltage amplifier (Trek, Model 5/80). The upper electrode was kept at ground potential. The liquid was quasi-statically injected with a syringe pump (KD Scientific, Legato 210 Series) through the feeding capillary. The droplet volume $\hat{\cal V}$ increased until its critical value was reached. The experiments verified the requirements $\rho g \hat{\cal V}^{2/3}/\gamma\lesssim 0.1$ and $H\gtrsim 3\hat{\cal V}^{1/3}$ for the critical volume not to be influenced by the gravity and the distance $H$ (provided that $E_0=V/H$ is fixed), respectively \cite{BHB19}. We decreased one order of magnitude the injected flow rate and checked that the results were practically the same. 

Digital images of droplet apex were taken using a high-speed video camera ({\sc Fastcam SA5}) equipped with optical lenses (12X {\sc Navitar}) and a microscope objective (10X {\sc Mitutoyo}). We processed the images with a subpixel resolution technique \citep{FLHMA13} to determine the interface contour at each instant. The apex curvature was calculated by fitting a Gaussian function to the interface contour in that region \citep{CE06}. 

The density $\rho$, viscosity $\mu$, surface tension $\gamma$, electrical conductivity $K$, and relative permittivity $\varepsilon$ of the working liquids are displayed in Table \ref{tab1}. The surface tension and electrical conductivity are sensitive to the presence of impurities. For this reason, it is advisable to measure their values. The surface tension was measured with the TIFA method \citep{CBMN06}. The electrical conductivity was determined by applying a voltage difference between the ends of a capillary filled with the liquid and measuring the resulting electric current. The slope of the straight line relating the applied voltage and electric current provides the conductivity value. The values of the rest of the liquid properties were taken from the manufacturer's specifications. 

\begin{table*}
\begin{center}
\begin{tabular}{cccccccc}
\hline 
&$\rho$ (kg/m$^3$) & $\mu$ (Pa s) & $\gamma$ (mN/m) & $K$ ($\mu$S/m)& $\varepsilon$ & $\delta_{\mu}$&$\alpha$\\
\hline
Glycerine& 1261 & 1.2 & 62.2 & 1.0 & 42.5 & $0.0292$& 6.05\\
\hline
\hline
1-Octanol& 827 & 0.0072 & 23.5 & 0.9 & 10 & 2.29 & 30.5\\
\hline
Deionized water& 997 & 0.001 & 71.0 & 54 & 80 & $9.36$ & 145\\
\hline
Triethylene glycol & 1113 & 0.021 & 44.2 & 24.1 & 41.4 & 0.44 & 167\\
\hline
Ethanol & 789 & 0.0011 & 22.1 & 3002 & 24.3 & 0.95 &481\\
\hline
Glycerine+LiCl 0.01M& 1261 & 1.2 & 62.2 & 192 & 42.5 &$0.00507$& 1163\\
\hline
\end{tabular}
\end{center}
\caption{Density $\rho$, viscosity $\mu$, surface tension $\gamma$, electrical conductivity $K$, and relative permittivity $\varepsilon$ of the working liquids at 25 $^\circ$C. The last two columns show the value of the electrohydrodynamic Reynolds number $\delta_\mu=[\gamma^2\rho\varepsilon_o/(\mu^3 K)]^{1/3}$ and dimensionless conductivity $\alpha=K\left[\rho R^3/(\gamma \varepsilon^2\varepsilon_o^2)\right]^{1/2}$. Glycerine+LiCl 0.01M corresponds to glycerine and LiCl dissolved at the molar concentration 0.01.} 
\label{tab1}
\end{table*}

Table \ref{tab1} also shows the values of the electrohydrodynamic Reynolds number $\delta_\mu=[\gamma^2\rho\varepsilon_o/(\mu^3 K)]^{1/3}$ and dimensionless conductivity $\alpha=K\left[\rho R^3/(\gamma \varepsilon^2\varepsilon_o^2)\right]^{1/2}$, where $\varepsilon_o$ is the vacuum permittivity.  

\subsection{Global stability analysis}

% Leaky-dielectric model
The leaky-dielectric model \citep{S97} is assumed to provide accurate results for small-amplitude perturbations around an equipotential base flow. This model is formulated in terms of the continuity and momentum equations,
\begin{equation}
\label{he}
\boldsymbol{\nabla}\cdot{\bf v}=0, \quad  \rho\frac{D{\bf v}}{Dt}=\rho {\bf g}-\boldsymbol{\nabla}p+\mu\boldsymbol{\nabla}^2 {\bf v},
\end{equation}
for the velocity ${\bf v}$ and pressure $p$ fields, where ${\bf g}$ is the gravitational acceleration. 

The hydrodynamic equations (\ref{he}) are integrated considering the kinematic compatibility condition and continuity of velocity and stresses at the interface. These stresses include the Maxwell stress
\begin{equation}
\label{Max}
{\boldsymbol \tau}_M^{(j)}=\varepsilon^{(j)} \left({\bf E}^{(j)}{\bf E}^{(j)}-\frac{1}{2}{\boldsymbol {\sf I}}E^{(j)2}\right),
\end{equation}
where $\varepsilon^{(j)}$ and ${\bf E}^{(j)}$ denote electrical permittivity and electric field evaluated on the phase $j$, respectively, while ${\boldsymbol {\sf I}}$ is the identity matrix. The dynamic effects of the outer gaseous medium are neglected. 

The Laplace equation for the electric potential is integrated considering the electric interface boundary conditions:
\begin{equation}
\label{maxs}
||\varepsilon^{(j)} {\bf E}^{(j)}||\cdot {\bf n}=\sigma_e, \quad ||{\bf E}^{(j)}||\times {\bf n}={\bf 0},
\end{equation}
where $||A||$ denotes the difference between $A$ evaluated at the two sides of the interface, ${\bf n}$ is the unit vector normal to the interface, and $\sigma_e$ is the surface charge density. The conservation equation for this quantity reads
\begin{equation}
\label{sigmae}
\frac{\partial \sigma_e}{\partial t}+{\boldsymbol \nabla}_s\cdot(\sigma_e {\bf v}_s)+\sigma_e ({\boldsymbol \nabla}_s\cdot {\bf n})({\bf v}\cdot {\bf n})=K E^{(i)}_n,
\end{equation}
where ${\boldsymbol \nabla}_s={\boldsymbol {\sf I}}_s {\boldsymbol \nabla}$ is the surface gradient operator, ${\boldsymbol {\sf I}}_s={\boldsymbol {\sf I}}-{\bf n}{\bf n})$ is the tensor that projects any vector onto the interface, and ${\bf v}_s={\boldsymbol {\sf I}}_s{\bf v}^{(j)}$ is the surface velocity, and $E^{(i)}_n$ the inner electric field normal to the interface. The formulation of the problem is completed by imposing the non-slip boundary condition at the solid surface, the voltages at the electrodes, the droplet volume, and the triple contact line anchorage condition.   

% Global stability analysis c
In the global stability analysis, one assumes the temporal dependence $\Phi=\Phi_0+\phi\, e^{-i\omega t}$, where $\Phi$ stands for any flow quantity, $\Phi_0$ is its value in the base flow, $\phi$ is the spatial dependence of the eigenmode, and $\omega=\omega_r+i\omega_i$ is the eigenfrequency. The droplet becomes unstable when the growth rate $\omega_i$ of the dominant eigenmode (i.e., that with the largest $\omega_i$)  becomes positive. 

% Numerical method
To solve the leaky-dielectric model, the inner and outer fluid domains are mapped onto two quadrangular domains through a non-singular mapping. The equations are discretized in the (mapped) radial direction with Chebyshev spectral collocation points \citep{KMA89}. We use fourth-order finite differences with equally spaced points to discretize the (mapped) axial direction. The method described by \citet{HM16a} allows one to obtain both the base flow and its eigenmodes. The calculation of the eigenmodes involves the Jacobian of the system evaluated with the base solution. The matrix accounting for the temporal dependence of the problem is calculated with essentially the same procedure as that for the Jacobian. More details of the leaky-dielectric model and the numerical method used to calculate the eigenmodes can be found in, e.g.,  Ref.\ \citep{PRHGM18}.

\section{Results}

\begin{figure}
\begin{center}
\includegraphics[width=0.8\linewidth]{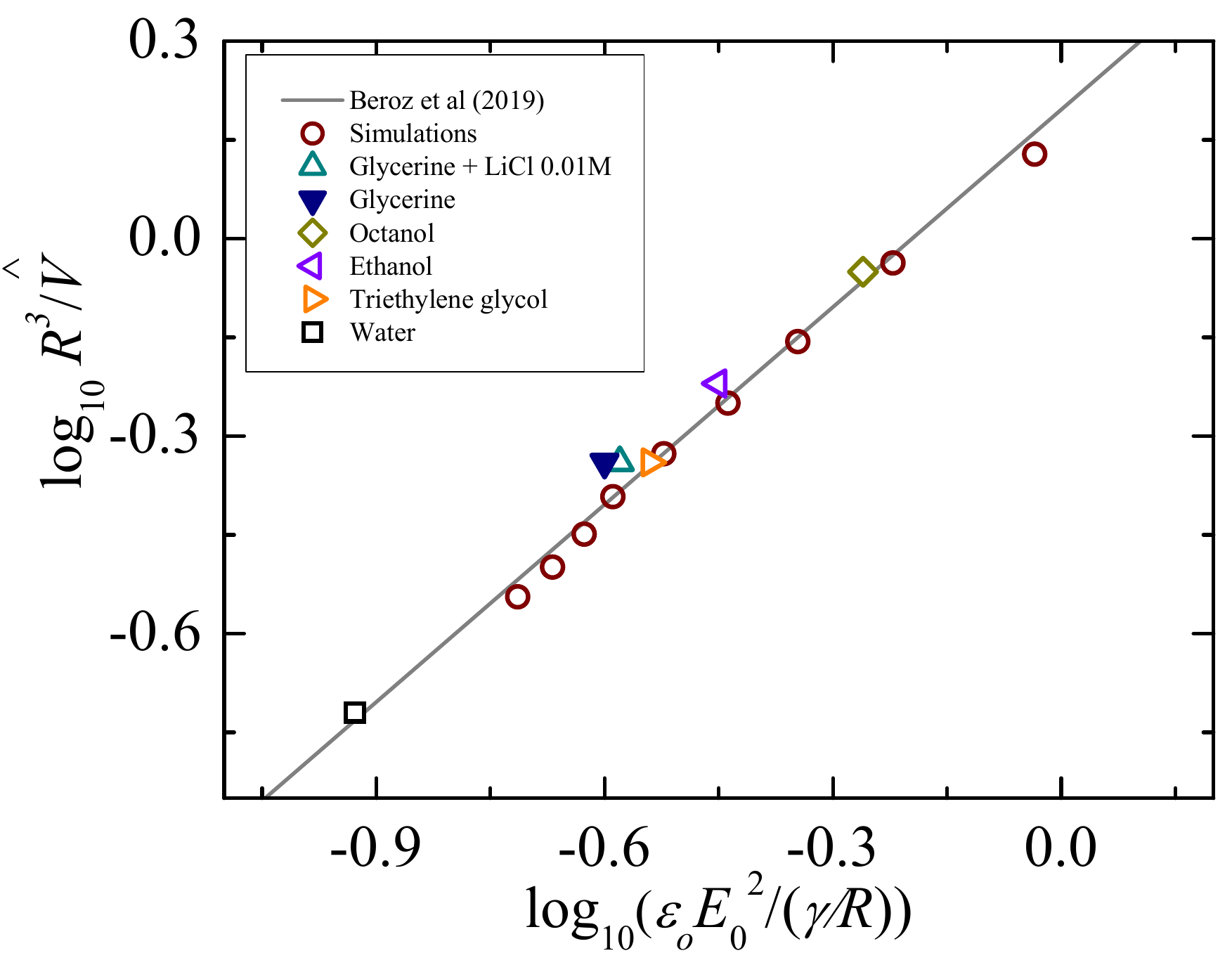}
\end{center}
\caption{Droplet volume at the onset of tip streaming as a function of the applied electric field $E_0=V/H$.}
\label{Beroz}
\end{figure}

\begin{figure}
\begin{center}
\includegraphics[width=\linewidth]{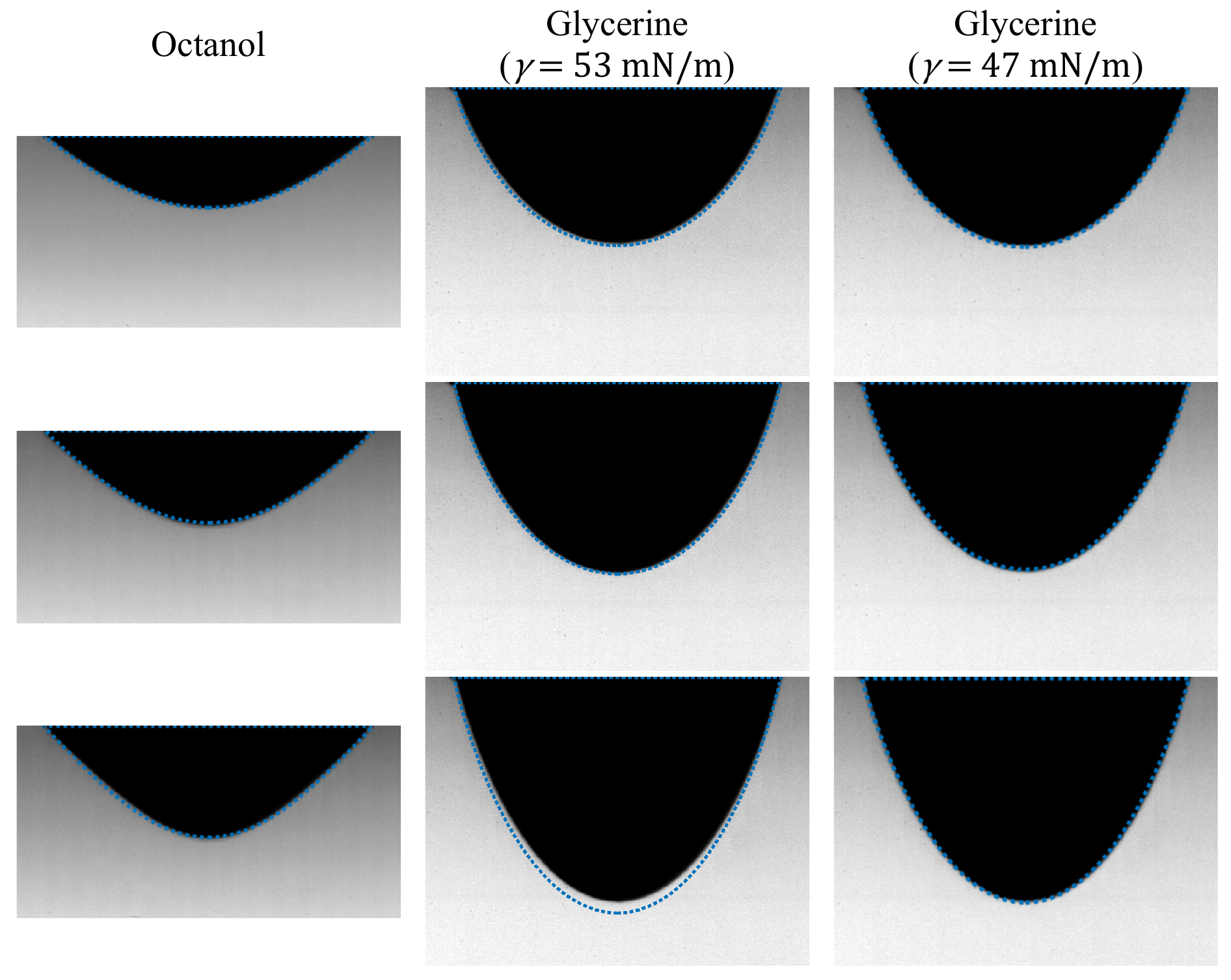}
\end{center}
\caption{Comparison between the experimental droplet shape and the numerical solution (dotted lines) for subcritical conditions.}
\label{shapes}
\end{figure}

% Comparision with Beroz
Figure \ref{Beroz} shows the agreement between the prediction of \citet{BHB19} and the results obtained from the global stability analysis. The figure also shows the critical volumes measured in our experiments. There is also a satisfactory agreement in practically all cases. We observe a slight discrepancy for glycerine and glycerine+LiCl 0.01M, which can be due to a slight influence of the electric field on the surface tension value. In fact, if we determine the value of $\gamma$ from the shapes of the electrified droplet (Fig.\ \ref{shapes}), the critical conditions coincide with the prediction of \citet{BHB19}. 

% Tip dynamics
We analyzed the breakup process for glycerine+LiCl 0.01M and glycerine, which have practically the same density, viscosity, surface tension, and permittivity, but very different electrical conductivities (see Table \ref{tab1}). We also examined the behavior of 1-octanol, whose electrical conductivity is similar to that of glycerine. 

\begin{figure}
\begin{center}
\includegraphics[width=0.8\linewidth]{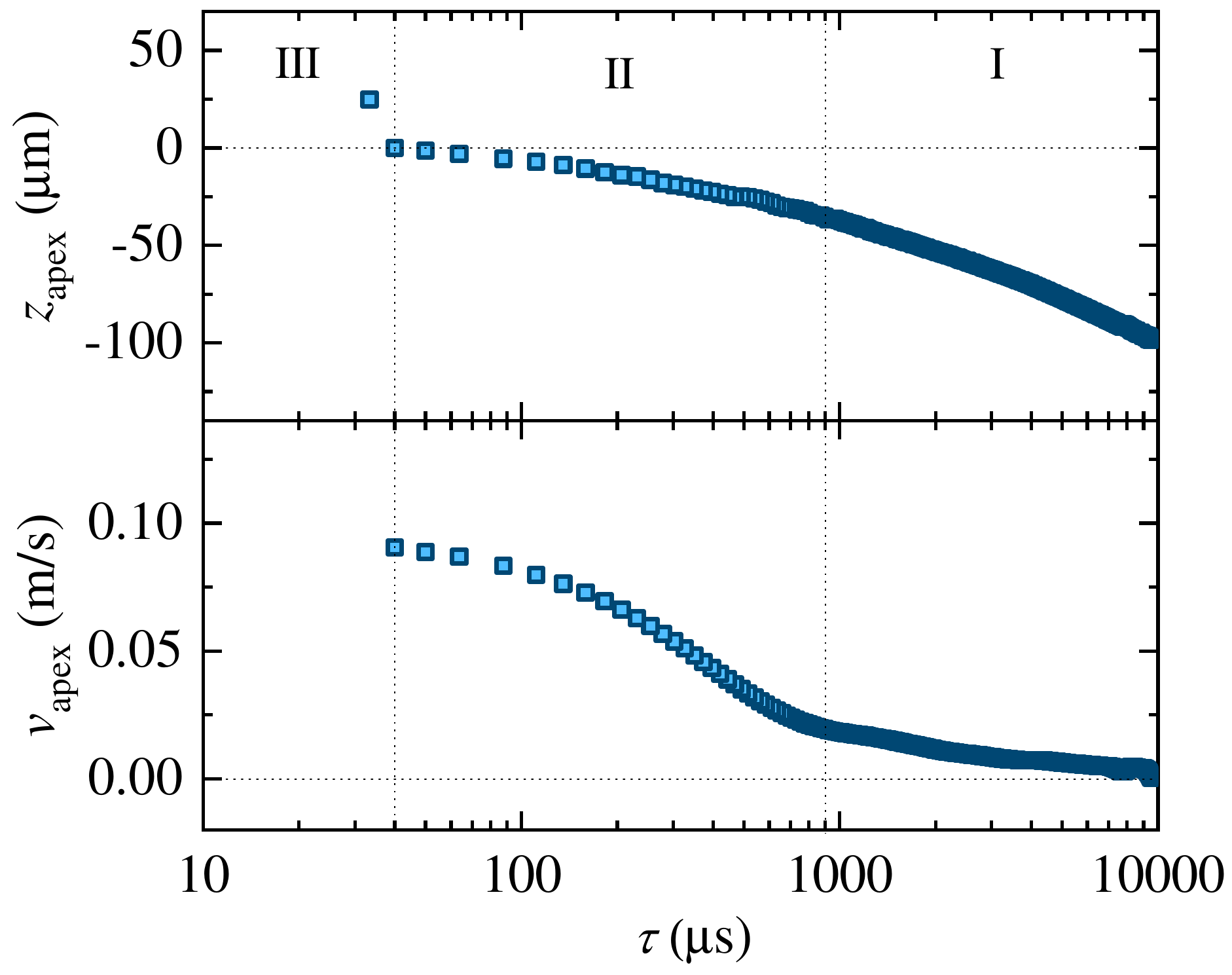}
\end{center}
\caption{Vertical coordinate $z_{\textin{apex}}$ and velocity $v_{\textin{apex}}$ of the apex of the glycerine$+$LiCl 0.01M droplet as a function of the time to the ejection $\tau$.}
\label{highconductivity}
\end{figure}

% The high-conductivity case
Figure \ref{highconductivity} shows the vertical coordinate $z_{\textin{apex}}$ and velocity $v_{\textin{apex}}$ of the apex of the glycerine$+$LiCl 0.01M droplet as a function of the time to the ejection $\tau$. Specifically, the time $\tau=0$ approximately corresponds to the instant at which the ejected liquid thread touches the counter-electrode for the first time. Therefore, the system advances in time as $\tau$ decreases. 

\begin{figure}
\begin{center}
\includegraphics[width=0.8\linewidth]{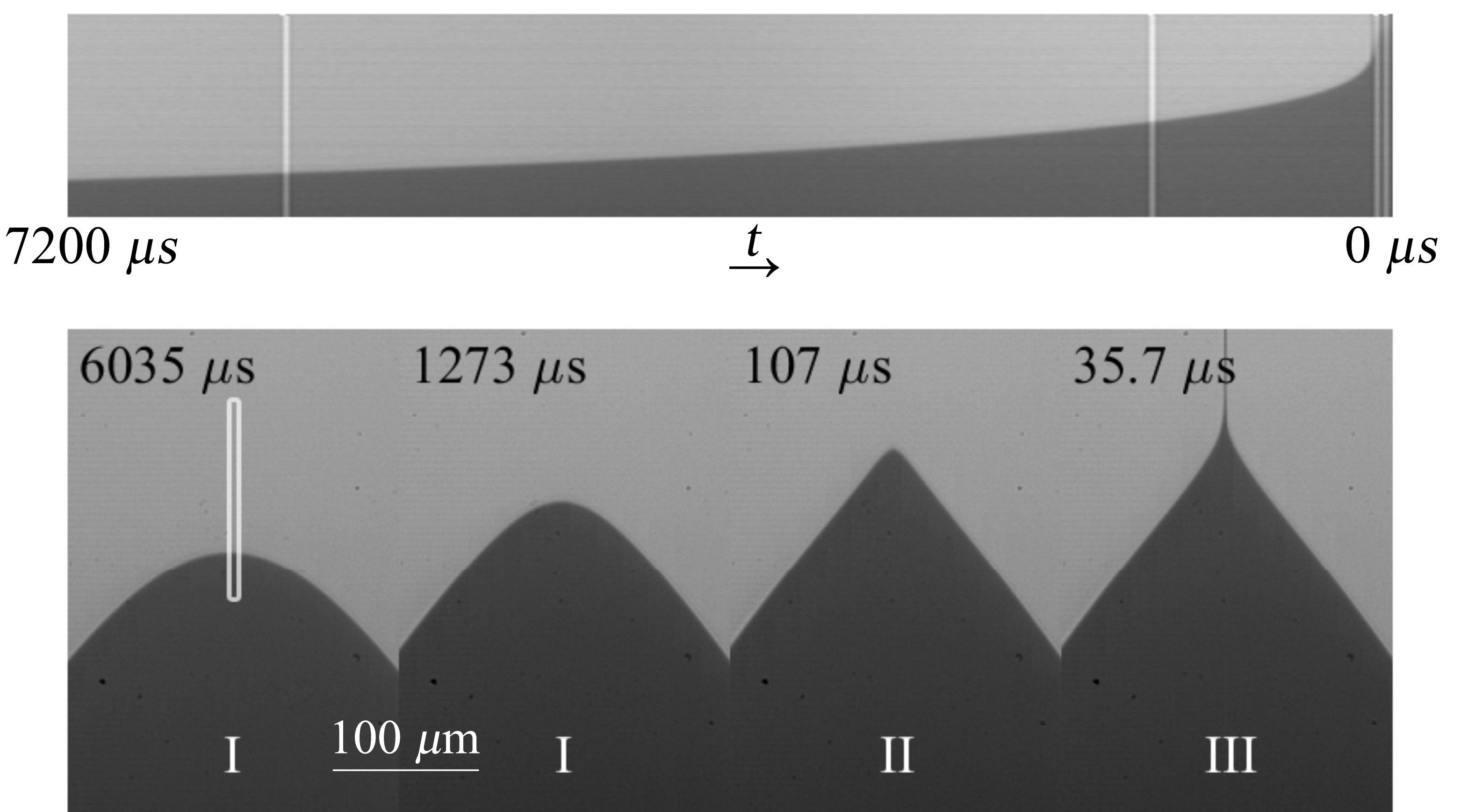}
\end{center}
\caption{Images of the apex of the glycerine$+$LiCl 0.01M droplet corresponding to the three phases shown in Fig.\ \ref{highconductivity}. The upper image shows the orthogonal projection of the pixel line coinciding with the droplet apex. The vertical white lines indicate the instants corresponding to the images shown below.}
\label{highconductivity2}
\end{figure}

The droplet of glycerine$+$LiCl 0.01M exhibited the behavior described by \citet{BHB19}. The apex moves very slowly towards the counter-electrode as the liquid is injected into the droplet [phase (I)]. Then, the droplet volume reaches the stability limit, and the apex stretches more rapidly [phase (II)]. This stretching gives rise to the jetting regime [phase (III)] in which a liquid thread is ejected. Hereafter, we choose $z_{\textin{apex}}=0$ as the apex vertical position at the transition to the jetting phase (III). We show the images of the droplet corresponding to these three phases in Fig.\ \ref{highconductivity2}.

% The low-conductivity case. 
The droplet behavior following the instability is drastically different in the pure glycerine case. This behavior depends on the characteristic electric field $V/H$ fixed during the experimental run. For sufficiently small values of $V/H$, the droplet does not emit a liquid thread regardless of the droplet volume $\hat{\cal V}$. For a critical value of $V/H$, the droplet apex oscillates before ejecting the liquid thread at the droplet volume threshold. The number of oscillations decreases as $V/H$ is increased. Finally, for sufficiently large values of $V/H$, the apex droplet ejects a liquid thread without previous oscillations, as occurs in the glycerine$+$LiCl 0.01M case. Here, we describe the experiment for $V=5$ kV and $H=3.3$ mm, in which the droplet apex oscillates once before the liquid ejection. 

\begin{figure}
\begin{center}
\includegraphics[width=0.8\linewidth]{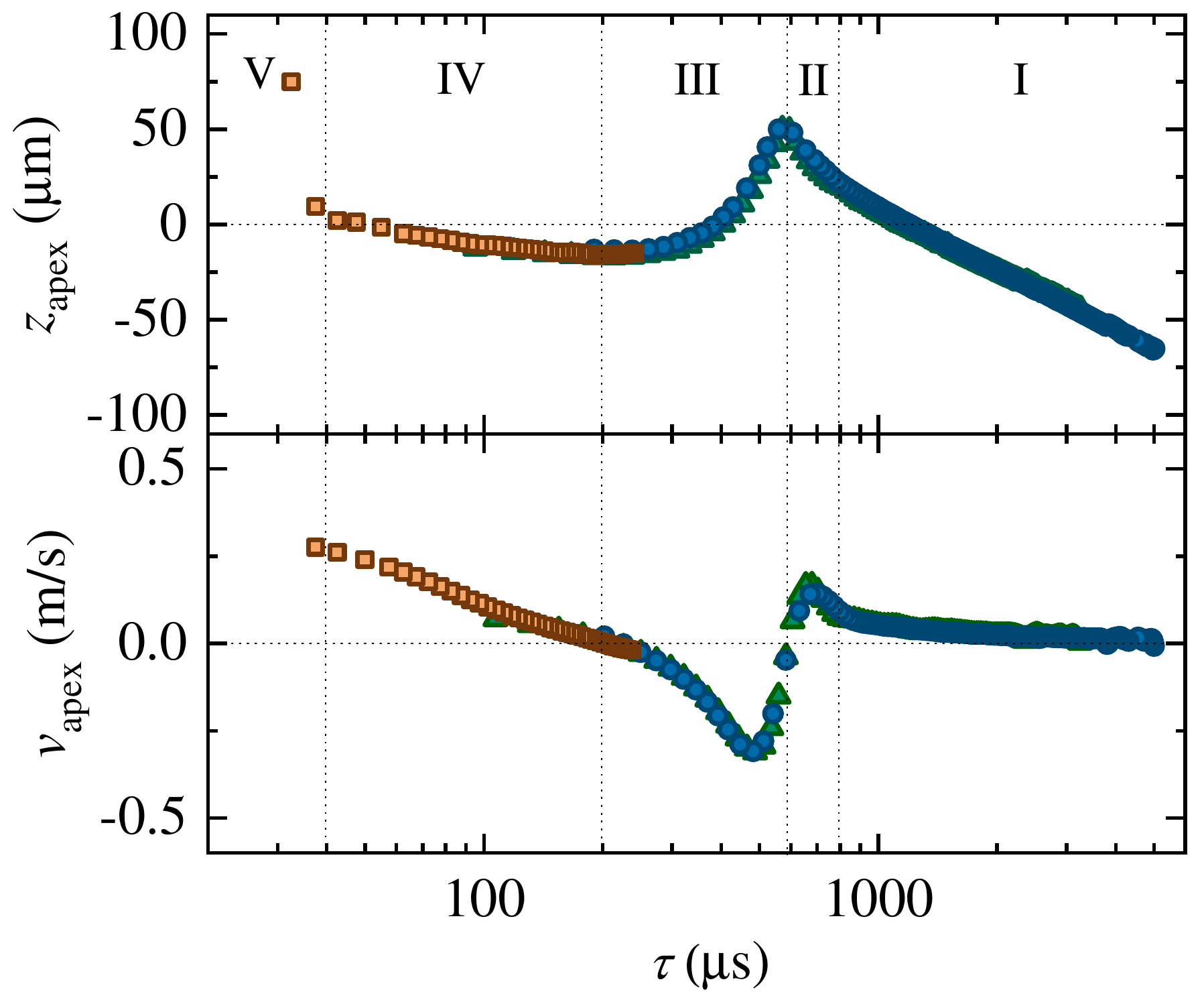}
\end{center}
\caption{Vertical coordinate $z_{\textin{apex}}$ and velocity $v_{\textin{apex}}$ of the apex of the glycerine droplet as a function of the time to the ejection $\tau$. The colors of the symbols correspond to different experiments.}
\label{lowconductivity}
\end{figure}

\begin{figure}
\begin{center}
\includegraphics[width=1\linewidth]{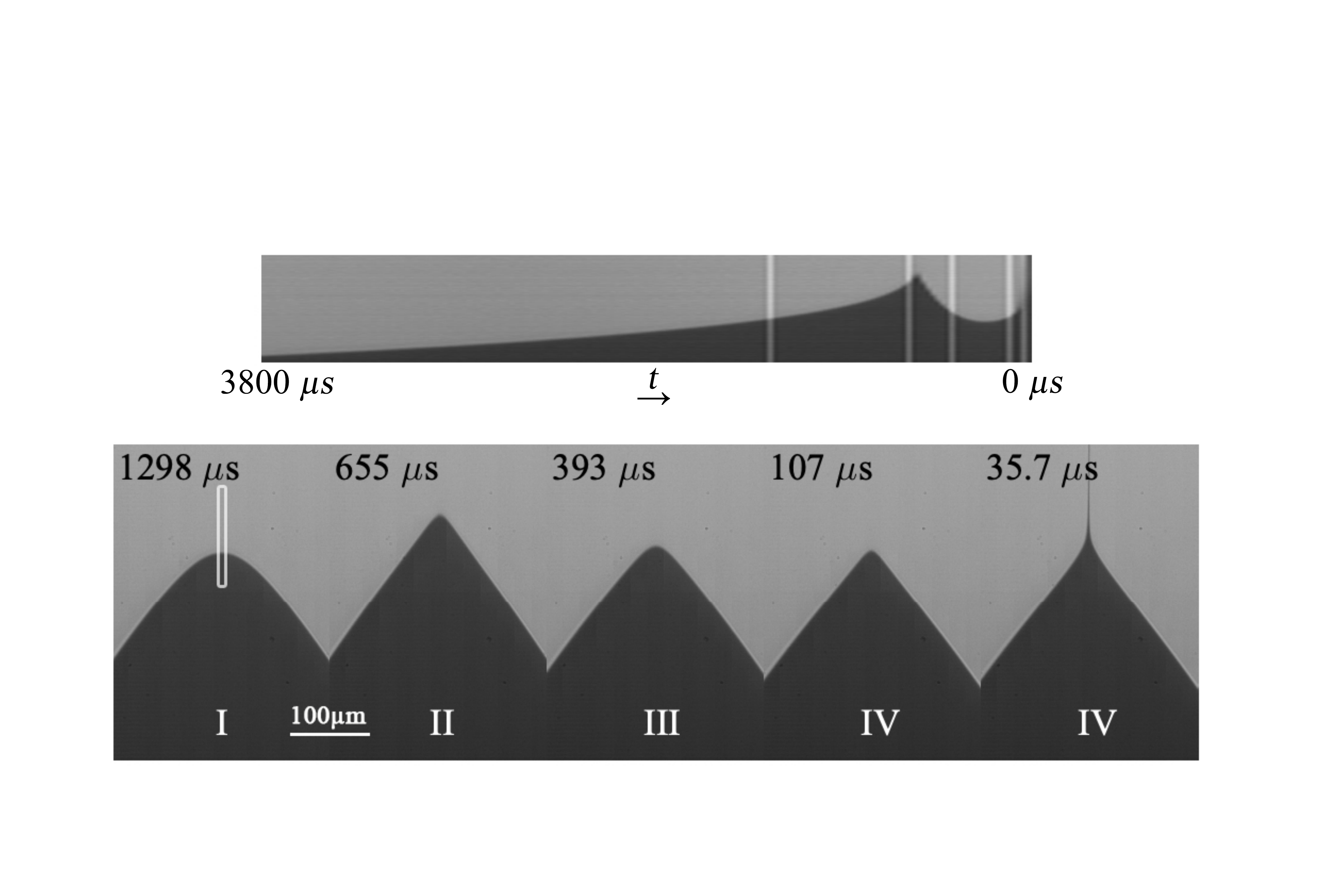}
\end{center}
\caption{Images of the apex of the glycerine droplet corresponding to the five phases shown in Fig.\ \ref{lowconductivity}. The upper image shows the orthogonal projection of the pixel line coinciding with the droplet apex. The vertical white lines indicate the instants corresponding to the images shown below.}
\label{lowconductivity2}
\end{figure}

\begin{figure}
\begin{center}
\includegraphics[width=0.8\linewidth]{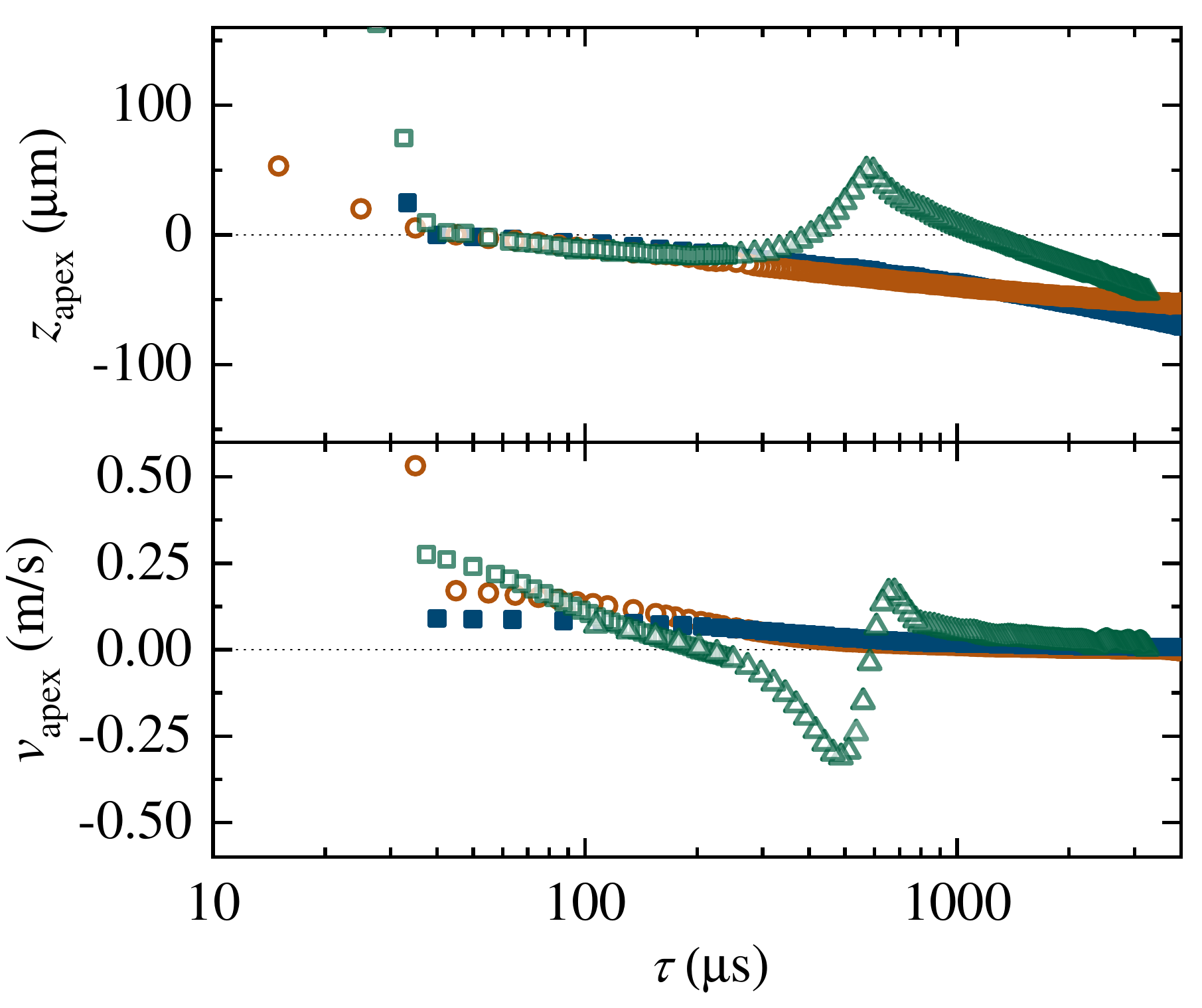}
\end{center}
\caption{Vertical coordinate $z_{\textin{apex}}$ and velocity $v_{\textin{apex}}$ of the apex of the glycerine$+$LiCl 0.01M (solid symbols), glycerine (green open symbols) and 1-octanol (orange open symbols) droplet as a function of the time to the ejection $\tau$.}
\label{comparision}
\end{figure}

% The one-oscillation case
Figure \ref{lowconductivity} shows the apex position and velocity corresponding to experiments conducted with different magnifications. The agreement among those results shows high experimental reproducibility. As in the previous case, the apex moves slowly towards the counter-electrode as the liquid is injected [phase (I)]. Then, the apex stretches more rapidly until the maximum value of $z_{\textin{apex}}$ is reached [phase (II)]. We have verified that there is no liquid ejection in this phase by increasing the magnification and the frame rate up to $10^6$ fps. In the next stage, the apex contracts [phase (III)] and bounces [phase (IV)]. This rebound finally leads to jet emission [phase (V)]. The images of the droplet corresponding to these five phases are shown in Fig.\ \ref{lowconductivity2}. The apex oscillation described above was not observed in the case of 1-octanol (Fig.\ \ref{comparision}), even though its conductivity is practically the same as that of glycerine.  

% The phenomenon
The glycerine droplet oscillation can be explained as follows. During the droplet quasi-static growth, the surface charge perfectly screens the inner electric field, and the liquid velocity vanishes. In this case, the electrostatic pressure is balanced by the capillary stress:
\begin{equation}
\label{balance}
\sigma_e E_n^{(o)}/2=\varepsilon^{(o)} E_n^{(o)2}/2=\gamma \hat{\kappa},
\end{equation}
where $E_n^{(o)}$ is the normal electric field evaluated on the outer side of the interface, and $\hat{\kappa}$ is the apex curvature. The surface charge density $\sigma_e$ and, therefore, the electrostatic pressure $\sigma_e E_n^{(o)}/2$ increase in the droplet apex as the volume increases. The curvature $\hat{\kappa}$ and, therefore, the capillary pressure $\gamma \hat{\kappa}$ also increase to balance the electrostatic pressure. This balance eventually breaks up, and the liquid moves towards the apex. 

The flow at this instability stretches the interface and increases the curvature at the apex. If the charges were supplied at the required rate, this process would follow a self-similar path \citep{GLRM16} in which the driver (electrostatic pressure) is balanced by inertia, surface tension and viscosity, according to the well-known temporal mechanisms of collapse \cite{E93}. However, the surface charge is not restored at the necessary rate in the glycerine case, and the surface charge density (electrostatic pressure) decreases while the capillary pressure increases. In this case, the electrostatic pressure is insufficient to overcome the resistant forces, and the apex retracts [phase (III)].

% Hypothesis
A natural question is why the glycerine droplet can eject a thin liquid thread after the oscillation. We hypothesize that the inner tangential electric field (which equals the inner one) gives rise to an electric shear stress that powers the liquid ejection after the apex rebound. The high liquid viscosity is a favorable factor in this phase because it ensures the transfer of momentum from the interface toward the bulk.

% Characteristic times
The above discussion suggests analyzing the phenomenon in terms of the inertio-capillary time $t_{ic}=(\rho R_i^3/\gamma)^{1/2}$ and the electric relaxation time $t_e=\varepsilon\varepsilon_o/K$. The ratio between these two quantities is the dimensionless electrical conductivity $\alpha=K\left[\rho R^3/(\gamma \varepsilon^2\varepsilon_o^2)\right]^{1/2}$. The value of $\alpha$ for glycerine is considerably smaller than for the rest of the liquids analyzed in this work (see Table \ref{tab1}). This indicates that charge is not transferred to the interface at the necessary speed in that case, which explains the anomalous behavior of this liquid. The dimensionless conductivity of octanol is much larger than that of glycerine, even though the two liquids have practically the same dimensional conductivity. For this reason, octanol does not exhibit charge relaxation effects (Fig.\ \ref{comparision}).  

\begin{figure}
\begin{center}
\includegraphics[width=0.8\linewidth]{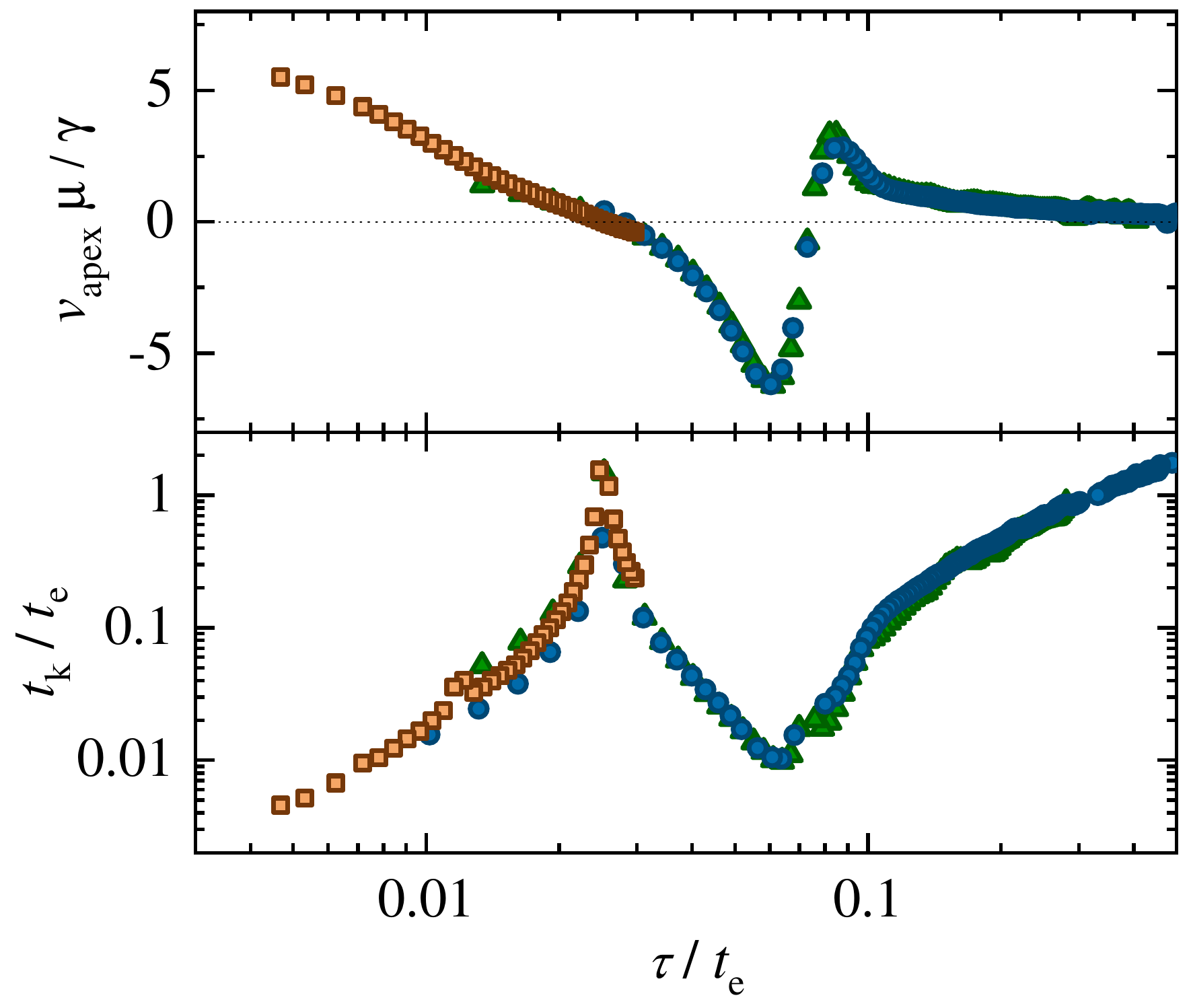}
\end{center}
\caption{(a) Vertical velocity $v_{\textin{apex}}$ in terms of the visco-capillary velocity $v_{\gamma\mu}\equiv \gamma/\mu$ as a function of $\tau/t_e$. (b) Kinematic time $t_k\equiv (v_{\textin{tip}} \hat{\kappa})^{-1}$ in terms of the electric relaxation time $t_e$ as a function of $\tau/t_e$. The liquid was glycerine. The colors of the symbols correspond to different experiments.}
\label{ele}
\end{figure}

% Other characteristic times
Figure \ref{ele} shows the values of $v_{\textin{apex}}/v_{\gamma\mu}$ ($v_{\gamma\mu}=\gamma/\mu$) for glycerine. As can be expected, the apex velocity takes values of the order of the viscous-capillary velocity $v_{\gamma\mu}$, showing the competition between the capillary driving force and the resistant viscosity force. The kinematic time $t_k=(v_{\textin{apex}} \hat{\kappa})^{-1}$ is a measure of the rate at which new interface is created by the flow, as indicated by the term $\sigma_e ({\boldsymbol \nabla}_s\cdot {\bf n})({\bf v}\cdot {\bf n})$ in Eq.\ (\ref{sigmae}). The results also show that $t_k/t_e\ll 1$ over most of the apex evolution, indicating that the interface production is much faster than the charge relaxation. The apex oscillation occurs in times much shorter than the electrical relaxation time ($\tau/t_e\ll 1$), which means that the electrical conductivity is insufficient to keep the electrostatic pressure necessary to trigger tip streaming. A significant inner electric field arises in the droplet apex due to the lack of charge relaxation. This electric field produces a shear stress responsible for the liquid ejection. 

% Global stability analysis
The tip curvature increases in one order of magnitude during the apex oscillation, which suggests that nonlinear effects become important in this process. We conducted a linear global stability analysis of this problem to confirm this hypothesis. As mentioned above, Fig.\ \ref{Beroz} shows the stability limit calculated from this analysis for arbitrary values of the applied voltage $V$, the distance $H$, and the liquid properties $\rho$, $\mu$, $\gamma$, $\varepsilon$, and $K$. The critical condition $\omega_i=0$ essentially depends on the droplet volume $\hat{{\cal V}}$, the ratio $V/H$, and the surface tension $\gamma$. The results perfectly agree with the prediction of \citet{BHB19}. 

\begin{figure}
\begin{center}
\includegraphics[width=0.8\linewidth]{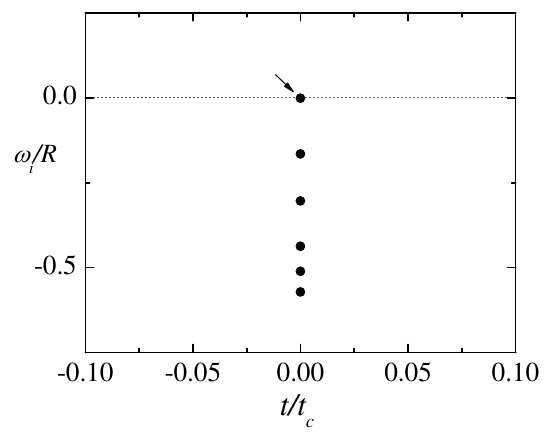}
\end{center}
\caption{Spectrum of eigenvalues with $\omega_i>-0.58$ for glycerine at the marginal stability. The arrow indicates the eigenvalue of the critical mode. The results for glycerine and glycerine$+$LiCl 0.01M overlap.}
\label{eigenvalues}
\end{figure}

We have analyzed the influence of the dimensionless conductivity $\alpha$ on the droplet linear dynamics at the marginal stability. For this purpose, the eigenmodes for glycerine and glycerine$+$LiCl 0.01M have been calculated. The eigenvalues (Fig.\ \ref{eigenvalues}) are practically the same in both cases. A non-oscillatory ($\omega_r=0$) instability arises for both glycerine and glycerine$+$LiCl 0.01M, which indicates that the linear analysis does not predict the oscillations of the glycerine droplet. 

\begin{figure}
\begin{center}
\includegraphics[width=0.8\linewidth]{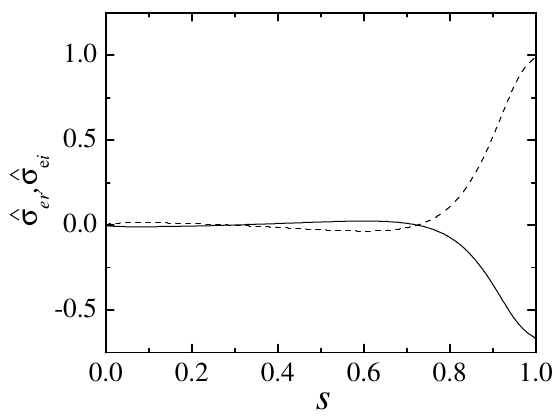}
\end{center}
\caption{Real $\hat{\sigma}_{er}$ (dashed line) and imaginary $\hat{\sigma}_{ei}$ (solid line) parts of the amplitude of the surface charge perturbation corresponding to the critical eigenmode at the marginal stability. The results are plotted versus the surface intrinsic coordinate $s$. The quantities $\hat{\sigma}_{er}$ and imaginary $\hat{\sigma}_{ei}$ have been made dimensionless by dividing them by $\left(\varepsilon_o \gamma/R\right)^{1/2}$. The surface intrinsic coordinate $s$ has been made dimensionless by dividing it by the value corresponding to the apex. The results for glycerine and glycerine$+$LiCl 0.01M overlap. The magnitude of the perturbation has been chosen arbitrarily.}
\label{charge}
\end{figure}

In the linear regime, the temporal evolution of the surface charge density is given by the expression 
\begin{equation}
\sigma_e=\sigma_{e0}+\text{Re}\left[(\hat{\sigma}_{er}+i\hat{\sigma}_{ei})e^{-i\omega t}\right],
\end{equation}
where $\sigma_{e0}$ is the charge distribution in the base flow, while $\hat{\sigma}_{er}$ and $\hat{\sigma}_{ei}$ are the real and imaginary parts of the eigenmode amplitude, respectively. This amplitude takes practically the same values for glycerine and glycerine$+$LiCl 0.01M. As shown in Fig.\ \ref{charge}, the amplitude of the surface charge perturbation increases in the apex. 

\begin{figure}
\begin{center}
\includegraphics[width=0.6\linewidth]{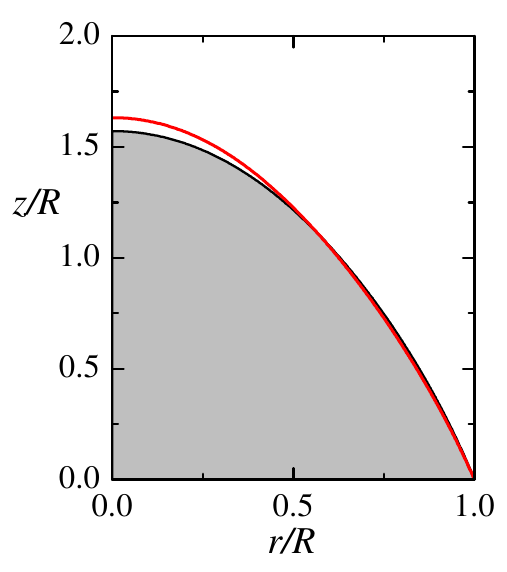}
\end{center}
\caption{Droplet shape in the base flow (shaded area) and interface displacement due to the critical eigenmode (red line) for glycerine at the marginal stability. The results for glycerine and glycerine$+$LiCl 0.01M overlap. The magnitude of the perturbation has been chosen arbitrarily to appreciate the interface deformation.}
\label{defor}
\end{figure}

Interestingly, the interface deformation caused by the critical eigenmode is not localized at the droplet tip (Fig.\ \ref{defor}), which means that the small local scale characterizing tip streaming is set during the nonlinear droplet deformation. Similar behavior has also been observed in a surfactant-loaded drop in an extensional flow \citep{HPREM22}. The interface deformations for glycerine and glycerine$+$LiCl 0.01M are also indistinguishable.   

In sum, and according to the leaky-dielectric model, the instability is not affected by the conductivity in the linear regime for the range of values considered in our analysis. The Ohmic conduction is sufficiently intense for Eq.\ (\ref{sigmae}) to be practically equivalent to $E^{(i)}_n=0$ (perfect conductor limit), and thus the conductivity value becomes irrelevant. In fact, we have verified that it is necessary to decrease the value of the dimensionless conductivity by several orders of magnitude for charge relaxation to be noticeable in the linear dynamics.  

\section{Concluding remarks}

The major finding of this work is the existence of oscillations prior to the liquid ejection in electrohydrodynamic tip streaming. This occurs when the dimensionless conductivity takes sufficiently small values. The droplet eventually emits a thin jet due to the force exerted by the electric shear stress arising during the oscillation. According to the leaky-dielectric model, the linear droplet dynamics is practically insensitive to the dimensionless conductivity for the range of values considered in our experiments. This indicates that the droplet oscillations result from the nonlinear evolution. In other words, charge relaxation effects appear only in the nonlinear phase of electrohydrodynamic tip streaming.

Numerical simulations have pointed out the role played by the electric shear stress during the jetting regime of electrified tip streaming [phase (III) in Fig.\ \ref{highconductivity}] \citep{CJHB08,FLHMA13}. In this case, the liquid overcomes an adverse pressure force beyond the base of the emitted jet driven by the electric shear stress exerted on the interface. Our experiments show that a similar mechanism is essential for the jet emission after the apex rebound in a droplet with low dimensionless conductivity, a phenomenon not observed previously.

The global stability analysis presented in this work is based on the leaky-dielectric model, which assumes Ohmic conduction with a constant electrical conductivity. One may wonder whether the linear eigenmode responsible for the instability accurately describes the small-amplitude dynamics in this problem. In fact, deviations from the leaky-dielectric model were found experimentally by \citet{GLRM16} during the formation of the first-emitted droplet. 

The phenomenon described above may be reproduced numerically if the appropriate physical model is considered, probably including electrokinetic effects. This study is beyond the scope of the present work.

\vspace{1cm}

Support from the Spanish Ministry of Science and Education (grant no. PID2019-108278RB-C32 / AEI / 10.13039/501100011033) and Gobierno de Extremadura (grant no. GR21091) is gratefully acknowledged. 

\bibliography{central}

\end{document}